\documentclass[conference]{IEEEtran}
\IEEEoverridecommandlockouts
\usepackage{comment}
\usepackage{cite}
\usepackage{amsmath,amssymb,amsfonts}
\usepackage{algorithmic}
\usepackage{graphicx}
\usepackage{subcaption}
\usepackage{textcomp}
\usepackage{xcolor}
\usepackage{stfloats}
\usepackage{url}

\def\BibTeX{{\rm B\kern-.05em{\sc i\kern-.025em b}\kern-.08em
    T\kern-.1667em\lower.7ex\hbox{E}\kern-.125emX}}

\begin{document}

\title{Resilience Through Escalation: A Graph-Based PACE Architecture for Satellite Threat Response}

\author{
  \IEEEauthorblockN{Anouar Boumeftah\IEEEauthorrefmark{1}, Sarah McKenzie-Picot\IEEEauthorrefmark{2}, Peter Klimas\IEEEauthorrefmark{2}, Gunes~Karabulut~Kurt\IEEEauthorrefmark{1}}
  \IEEEauthorblockA{\IEEEauthorrefmark{1}Poly-Grames Research Center, Department of Electrical Engineering, Polytechnique Montr\'{e}al, Montr\'{e}al, QC, Canada\\
  \IEEEauthorrefmark{2}NorthStar Earth \& Space, Montr\'{e}al, QC, Canada\\
  \textit{\{anouar.boumeftah, gunes.kurt\}@polymtl.ca, \{sarah.mckenzie-picot, peter.klimas\}@northstar-data.com}}
}

\maketitle

\begin{abstract}
Modern satellite systems face increasing operational risks from jamming, cyberattacks, and electromagnetic disruptions in contested space environments. Traditional redundancy strategies often fall short against such dynamic and multi-vector threats. This paper introduces a resilience-by-design framework grounded in the PACE methodology, which stands for Primary, Alternate, Contingency, and Emergency, originally developed for tactical communications in military operations. It adapts this framework to satellite systems through a layered state-transition model informed by threat scoring frameworks such as CVSS, DREAD, and NASA’s risk matrix. We define a dynamic resilience index to quantify system adaptability and implement three PACE variants  including static, adaptive, and \(\epsilon\)-greedy reward-optimized to evaluate resilience under diverse disruption scenarios. Results show that lightweight, decision-aware fallback mechanisms can substantially improve survivability and operational continuity for next-generation space assets.
\end{abstract}                                                                                                                                           


\section{Introduction}\label{sec:introduction}

Satellite systems are operating in an increasingly volatile threat environment, where disruptions may result not only from natural phenomena but also from deliberate interference. With the growing reliance on space-based infrastructure for communications, navigation, Earth observation, and defense operations, orbital assets have become increasingly vulnerable components of integrated terrestrial and non-terrestrial networks. These systems face a range of risks that include cyberattacks, jamming, signal spoofing, and electromagnetic disturbances, as well as cascading failures triggered by software or control anomalies across space-ground link segments. In particular, recent operational disruptions have demonstrated the potential of satellite interference to impact real-world missions. For example, the 2022 cyberattack on Viasat satellite terminals in Europe disrupted connectivity for both civilian and military users, underscoring the cross-domain consequences of digital vulnerabilities in space systems~\cite{Bingen2022}. Similarly, the 1962 Starfish Prime test remains a reference for the destructive reach of high-altitude electromagnetic pulse (EMP) events, which can cause systemic satellite outages~\cite{Lele2024}.

\subsection{Background and Related Work}\label{sec:background}

Resilience in satellite systems refers to the ability to sustain essential functions in the presence of both hardware faults and malicious threats. Recent research has emphasized that while space assets are increasingly targeted by cyberattacks, many systems still lack robust, integrated resilience strategies. Pavur \textit{et al.}~\cite{pavur2022} document decades of satellite hacking incidents, underscoring the vulnerability of the sector. Falco \textit{et al.}~\cite{falco2019} emphasize the importance of dynamic requirements based on mission objectives. Military and national space policies now stress resilience as a blend of fault tolerance and cyber defense~\cite{murat2023secure}, requiring satellites to withstand both stochastic failures and deliberate attacks. In engineering, redundancy remains a prime tactic, often implemented via $N+1$ schemes where $N$ represents the minimum number of components required for normal operation, and the +1 denotes an additional redundant component, or triple modular redundancy (TMR). However, these methods are largely static, and while effective against random faults, they may fail when confronted with coordinated or adaptive attacks. Sun \textit{et al.}~\cite{sun2022bayesian} apply Bayesian methods to assess satellite redundancy effectiveness, and O'Halloran \textit{et al.}~\cite{ohalloran2017} present a framework for analyzing failure propagation in cyber-physical systems during early design stages, emphasizing the identification of forward, backward, and uncoupled propagation paths using graph-theoretic metrics. Xia \textit{et al.}~\cite{xia2019} extend this with a directed graph approach for complex electromechanical systems (CMESs) under attack. From a system-level view, the MILSATCOM DFARR initiative offers insights into designing affordable, resilient SATCOM systems~\cite{glaser2013dfarr}. Its focus on modular waveforms and flexible ground-space processing reflects a systems engineering approach to cyber and jamming resilience, aligning with emerging policies promoting layered protection against kinetic, electronic, and software-based threats.

\subsection{Existing Threat Modeling Frameworks}\label{sec:existing_threat_frameworks}

A Threat Model is a formal abstraction that identifies potential adversarial and non-adversarial hazards, system vulnerabilities, and the pathways through which threats can be realized. It supports risk-informed decision-making across system design, deployment, and contingency planning. NASA's risk matrix classifies hazards by likelihood and severity, while cybersecurity frameworks such as CVSS and DREAD assess vulnerabilities and threats. CVSS provides quantitative software risk scores, whereas DREAD qualitatively ranks threats based on factors like damage potential and exploitability.

For satellite systems, these models offer useful insights but are not yet well integrated into onboard decision-making. Pavur \textit{et al.}~\cite{pavur2022} and Falco \textit{et al.}~\cite{falco2019} advocate for unified threat modeling that bridges space-specific hazards with cyber risk assessment. Similarly, Jacobsen and Marandi propose a STRIDE-based cyber-threat taxonomy for unmanned aerial systems, identifying vulnerabilities such as spoofing and privilege escalation ~\cite{jacobsen2021uav}. Their framework is transferable to satellites, especially as autonomous platforms increase cyber-physical integration risks.

Despite their value, existing threat models are often applied in isolation and remain disconnected from onboard decision-making systems. This limits their utility in dynamic, contested environments that demand rapid adaptation. To overcome this gap, Section~\ref{sec:proposed_framework} presents a structured, decision-aware fallback framework designed to support resilient satellite operations under diverse threat conditions.

\subsection{Origins of the PACE Methodology}\label{sec:pace_overview}

The PACE model, denoting \textit{Primary}, \textit{Alternate}, \textit{Contingency}, and \textit{Emergency}, originated in U.S. Special Operations Forces as a structured planning framework to maintain mission continuity in degraded, denied, or contested environments~\cite{army2023fm602}. It assumes that a single backup is insufficient in high-risk contexts, instead promoting a hierarchical set of functionally independent alternatives ranked by operational feasibility and reliability. Originally applied to tactical communications, PACE has since been adopted across other high-reliability domains. In military command and control, each level represents an autonomous method for achieving a critical function, supporting resilience against system failure or adversarial disruption~\cite{army2023health}. Emergency medical systems apply the model to sequence critical interventions, enabling clinicians to transition rapidly between escalating response protocols under time constraints~\cite{winniford2025pace}. Similarly, agencies such as the Cybersecurity and Infrastructure Security Agency (CISA) employ PACE to architect redundant communication channels for national infrastructure and emergency response operations~\cite{cisa2023pace}.

\subsection{Research Gap and Contributions}\label{sec:research_gap}

Classical redundancy techniques improve fault tolerance but are static and optimized for random, independent failures. In contested environments, they fail to address adaptive or coordinated disruptions, often resulting in common mode failures where redundant units share the same fate~\cite{sterbenz2010resilience,jacobs2012reconfigurable}. These designs also impose significant overhead without ensuring resilience against evolving threats. Recent works emphasize the need for context-aware, adaptive redundancy. Jacobs~\textit{et al.} propose reconfigurable fault tolerance that scales with environmental stress~\cite{jacobs2012reconfigurable}. Buys~\textit{et al.} show static schemes underperform in dynamic settings and advocate adaptive management~\cite{buys2011adaptive}. Ahmed and Bhargava introduce replica reconfiguration to counter adversarial disruptions~\cite{ahmed2018byzantine}. George~\textit{et al.} demonstrate multimode spacecraft architectures that switch between TMR and symmetric multiprocessing~\cite{george2017hybrid}. Graph-based methods, such as knowledge graphs for anomaly detection, further support context-aware system-level resilience~\cite{yi2023knowledge,xia2019}. Together, these studies highlight the need for adaptive, multi-layered strategies. To address these challenges, we investigate the use of a layered architecture inspired by the PACE planning concept. The key contributions of this work are as follows:

\begin{itemize}
    \item Adaptation of the PACE resilience framework, originally developed in military communications, for satellite operations in contested orbital environments.
    \item Design of a threat-driven decision model that selects optimal PACE fallback layers based on mission utility and operational cost.
    \item Formalization of dynamic transitions using an \(\epsilon\)-greedy, reward-driven mechanism to enable traceable escalation and recovery strategies.
    \item A system-level model integrating threat likelihood, operational cost, and utility, evaluated via Monte Carlo simulations of a jamming scenario.
    \item Definition of the dynamic redundancy efficiency index (DREI) to quantify the trade-off between mission performance and fallback cost across strategies.
\end{itemize}

The remainder of this paper is organized as follows. Section~\ref{sec:threats_vulnerabilities} presents the system model and threat scenarios. Section~\ref{sec:proposed_framework} introduces the graph-based PACE framework and its transition logic. Section~\ref{sec:resilience_metrics} defines DREI. Section~\ref{sec:simulation_setup} outlines the simulation setup and PACE variants. Section~\ref{sec:results_discussion} analyzes the results. Section~\ref{sec:conclusion} concludes the paper.

\section{System and Threat Overview}\label{sec:threats_vulnerabilities}

\begin{figure*}[!t]
  \centering
  \includegraphics[width=0.92\textwidth]{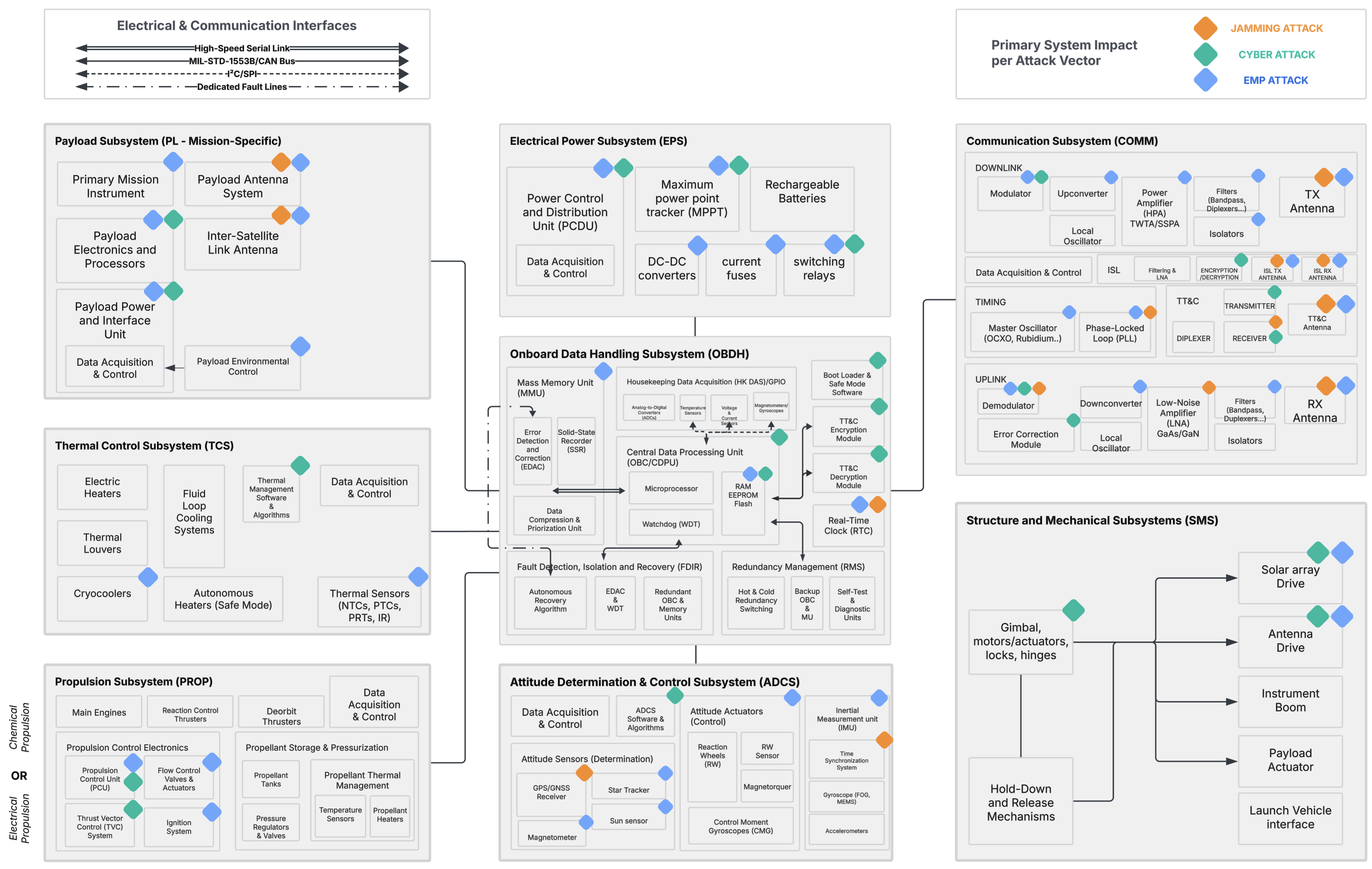}
  \caption{System-level block diagram of a representative satellite architecture. Each large block denotes a major subsystem. Colored diamond markers indicate primary vulnerability by attack vector: jamming (orange), cyber (green), and electromagnetic pulse (blue).}
  \label{fig:satellite_system_diagram}
\end{figure*}

\subsection{Threat Taxonomy Across Space Segments}\label{sec:satellite_segments}

Threats to satellite systems can be categorized across three operational domains: the \textit{space}, \textit{ground}, and \textit{link} segments. While this segmentation helps targeted risk assessment, many threat vectors span across these domains, leading to interdependent or cascading system failures as discussed in Section~\ref{sec:cascading_failures}.

\subsubsection{Space Segment}\label{sec:space_segment}

The space segment includes the onboard systems presented in Fig.~\ref{fig:satellite_system_diagram}, which are vulnerable to both hardware and software threats. Willbold \textit{et al.}~identify multiple security-critical flaws in satellite firmware, including lack of access controls and unprotected telecommand interfaces, which may enable persistent malware or even full system takeover~\cite{willbold2023spaceodyssey}. Additional risks stem from unvetted commercial off-the-shelf (COTS) components and firmware backdoors introduced through supply chains, as emphasized in Falco's taxonomy~\cite{falco2021taxonomy}.

\subsubsection{Ground Segment}\label{sec:ground_segment}

The ground segment covers mission control, antenna networks, and supporting cloud infrastructure. Common threats include poor network segmentation, outdated telemetry protocols, and misconfigured APIs~\cite{enisa2025}. Attacks may originate from compromised remote portals and escalate to unauthorized command injection. Past incidents, including the ROSAT and KA-SAT breaches, highlight how weak patching and insecure interfaces have caused full system disruptions~\cite{falco2021taxonomy}.

\subsubsection{Link Segment}\label{sec:link_segment}

The link segment includes satellite-ground communication and inter-satellite links (ISLs). It is highly exposed to RF-based threats such as jamming, spoofing, and eavesdropping. Baselt \textit{et al.}~report widespread transmission of unencrypted SATCOM data~\cite{baselt2022satcom}, while Morales-Ferre \textit{et al.}~show how global navigation satellite system (GNSS) spoofing can corrupt satellite timing and positioning~\cite{morales2020gnss}. These attacks can force degraded modes, induce orbital errors, or enable adversarial control via uplink spoofing~\cite{enisa2025}.

\subsection{System-Level Modeling and Impact Taxonomy}\label{sec:major_threats}

The system-level threat mapping in Fig.~\ref{fig:satellite_system_diagram} illustrates how key subsystems are differently exposed to jamming, cyber, and EMP attacks. We focus our analysis on three representative threat classes with significant cross-segment impact: signal jamming and spoofing, cyberattacks, and EMP events. The numerical evaluation in Section~\ref{sec:results_discussion} addresses jamming scenarios.

\subsubsection{Signal Jamming and Spoofing}\label{sec:jamming_spoofing}

Jamming disrupts uplink and downlink channels, while spoofing injects deceptive signals into control loops. ENISA documents how such disruptions can trigger fallback modes or delay operations~\cite{enisa2025}. Morales-Ferre~\textit{et al.} analyze replay and meaconing attacks on GNSS, showing the feasibility of corrupting orbital data and synchronization in assets reliant on satellite navigation~\cite{morales2020gnss}. These threats are increasingly relevant for space assets using GNSS for orbit maintenance or timing. Adaptive detection methods for on-orbit jamming and proximity threats have also been proposed to complement such resilience frameworks~\cite{anouar1,anouar3}.

\subsubsection{Cyberattacks}\label{sec:cyberattacks}

Cyber intrusions target satellite software stacks, including command units, firmware, and communication protocols. Willbold~\textit{et al.} reveal vulnerabilities that allow persistent footholds across subsystems~\cite{willbold2023spaceodyssey}. ENISA classifies digital threats across the satellite lifecycle, citing development-time vulnerabilities, unencrypted control links, and weak in-orbit cryptography~\cite{enisa2025}. Deloitte highlights growing risks from satellite interconnectivity and the adoption of ground station as a service (GSaaS), which lowers barriers for adversaries~\cite{deloitte2024}.

\subsubsection{Electromagnetic Pulse (EMP)}\label{sec:emp}

EMP events can impact all segments simultaneously. Kopp describes how nuclear and non-nuclear EMPs induce destructive surges across conductors, damaging power systems, front ends, and avionics~\cite{kopp1996emp}. Satellites using COTS microelectronics are particularly vulnerable to latch-ups and burnout. ENISA lists EMP among the top non-kinetic threats due to its systemic and instantaneous damage potential~\cite{enisa2025}.

\subsection{Cascading Failure Effects Across Subsystems}\label{sec:cascading_failures}

Attacks in one segment often trigger system-wide degradation. Spoofed links can mislead on-board computers (OBC), and ground-based cyber intrusions may corrupt firmware and disable safeguards. Falco's study shows that compromised satellites can spread malware across constellations via ISLs~\cite{falco2020whensatellitesattack}. EMPs can simultaneously disable telemetry, power, and attitude control. These cascading effects underscore the need for resilient fallback architectures that isolate faults and preserve critical functionality, which is a core goal of the PACE-based framework.

The threats outlined above motivate a unified resilience framework. While models like NASA's risk matrix, CVSS, and DREAD offer valuable assessments, they are not interoperable and lack integration with real-time decision-making. Section~\ref{sec:proposed_framework} presents a layered, decision-aware architecture that combines threat modeling, operational cost, and mission utility to support adaptive fallback strategies.

\section{Proposed Framework}\label{sec:proposed_framework}

\subsection{PACE-Based Resilience Model}\label{sec:layered_model}

This model organizes operational behavior into four resilience layers: \textbf{Primary} (nominal), \textbf{Alternate} (degraded but functional), \textbf{Contingency} (survivability under stress), and \textbf{Emergency} (minimal-operation or distress signaling). Each layer reflects a trade-off between mission utility and system burden, balancing performance, risk tolerance, and resource consumption. Together, these layers form a modular, adaptive structure for graceful degradation. Transitions are guided by context-aware metrics and pre-defined thresholds to maintain continuity in uncertain environments~\cite{army2023health}.

\subsection{Graph-Based Representation}\label{sec:graph_representation}

\begin{figure*}[!t]
  \centering
  \includegraphics[width=0.85\textwidth]{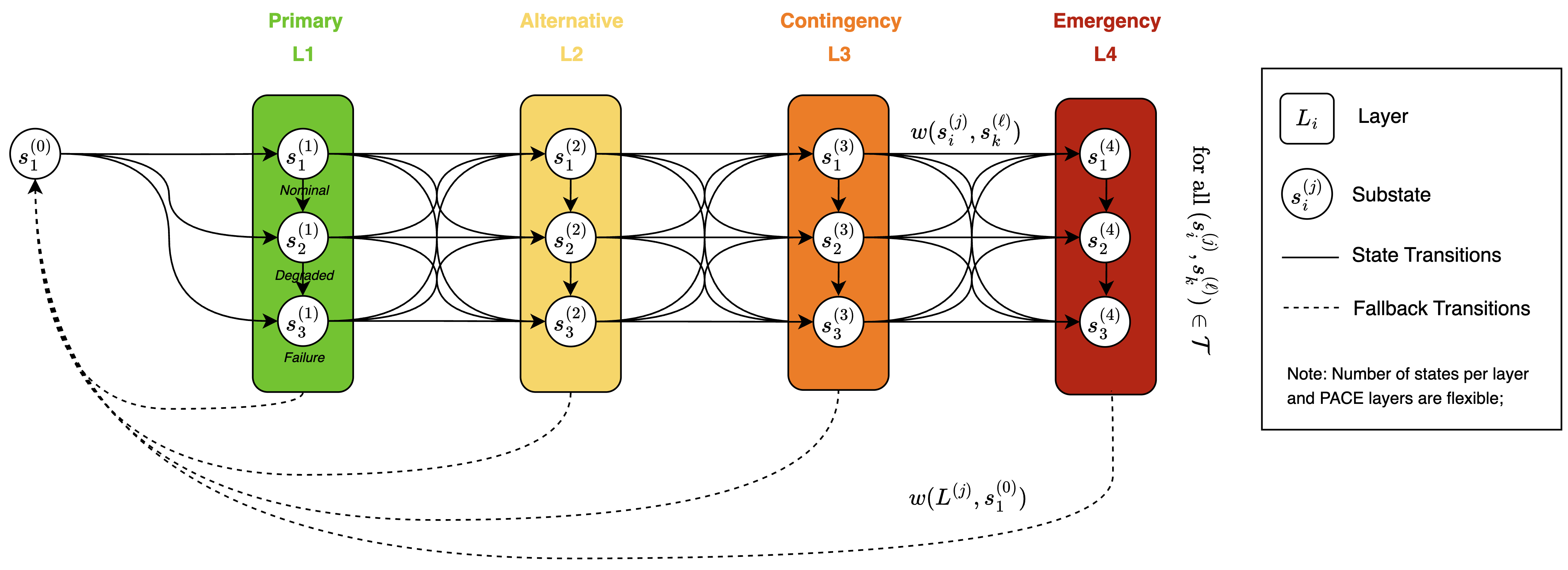}
  \caption{PACE graph-based representation diagram.}
  \label{fig:pace_main_diagram}
\end{figure*}

As shown in Fig.~\ref{fig:pace_main_diagram}, we formalize the PACE-based framework as a directed multi-layer graph \(G = (\mathcal{S}, \mathcal{T})\), where \(\mathcal{S} = \bigcup_{i=1}^4 \{s_i^{(j)}\}\) is the set of system states across all layers \(L_i \in \{\text{P}, \text{A}, \text{C}, \text{E}\}\), and \(\mathcal{T} \subseteq \mathcal{S} \times \mathcal{S}\) is the set of transitions between states.

Each state \(s \in \mathcal{S}\) is associated with a utility function \(\omega(s)\) and a cost-adjusted recovery potential. Transitions \((s, s') \in \mathcal{T}\) are annotated with:
\begin{equation}
w(s, s') = \left(p(s, s'),\; c(s, s')\right)
\end{equation}
where \(p(s, s')\) is the transition probability and \(c(s, s')\) is the associated cost (e.g., energy, reconfiguration time, utility drop).

\subsubsection{State Definition and Layering}\label{sec:state_definition}

Each layer \(L_i\) contains a fixed number of states \(s_i^{(j)}\), representing degrees of operational viability. Horizontal transitions handle degradation or recovery within the same layer; vertical transitions represent fallback under threat or upward recovery, subject to cost and timing constraints.

\subsubsection{Transition Probability and Cost Modeling}\label{sec:transition_modeling}

Transition weights are modeled jointly based on threat likelihood (e.g., NASA risk matrices, CVSS scores), operational cost (e.g., resource use, mission loss, delays), and system dynamics that adapt to environmental factors such as jamming intensity, energy reserves, and mission phase.

The threat mapping is as follows: let \(\rho \in [0,1]\) denote a normalized threat score. For CVSS, \(\rho = \text{CVSS}/10\); for NASA likelihood-severity bins, \(\rho\) is read from a predefined \(5 \times 5\) lookup table with entries \(\rho(\ell, \sigma) = \ell \cdot \sigma / 25\), where \(\ell\) and \(\sigma\) denote likelihood and severity levels respectively. A base downward transition probability from layer \(L_i\) to \(L_{i+1}\) is then:
\begin{equation}
p_{\text{base}}(L_i \to L_{i+1}) = \rho \cdot p_{\max},
\end{equation}
where \(p_{\max} \in (0,1)\) caps the worst-case transition rate. Base costs are assigned proportionally to utility loss between layers. The adaptive and \(\epsilon\)-greedy variants modulate these base values at runtime as described in Section~\ref{sec:pace_variants}.

\section{Resilience Evaluation Index}\label{sec:resilience_metrics}

To quantify how effectively a satellite system maintains operational continuity under disruption, we introduce the \textit{Dynamic Redundancy Efficiency Index (DREI)}, which captures the trade-off between mission utility and the cost of fallback or recovery transitions over time. Let \(\mathcal{S}\) denote the set of all system states, and \(P_t(s)\) the probability of being in state \(s \in \mathcal{S}\) at time \(t\). Each state \(s\) is associated with a utility value \(\omega(s)\), and transitions \((s, s')\) incur a cost \(c_t(s, s')\) at time \(t\).

The adjusted utility \(\omega_t^*(s)\) is defined as:
\begin{equation}
\omega_t^*(s) =
\begin{cases}
\omega(s) \cdot \kappa, & \text{if } s = \text{P\_Nominal and } t > 0 \\
\omega(s), & \text{otherwise,}
\end{cases}
\end{equation}
where \(\kappa > 1\) is a resilience reward multiplier that emphasizes the strategic value of returning to nominal operation. The DREI at time \(t\) is then:
\begin{equation}
\text{DREI}_t = \frac{\displaystyle\sum_{s \in \mathcal{S}} \omega_t^*(s) \cdot P_t(s)}{C_t},
\end{equation}
where \(C_t = \sum_{\tau=1}^{t} \sum_{(s,s') \in \mathcal{T}} c_\tau(s,s') \cdot \mathbf{1}[\text{transition taken at } \tau]\) is the total accumulated transition cost up to time \(t\). The multiplier is fixed at \(\kappa = 1.2\) throughout all simulations, reflecting the modest but meaningful strategic value of recovering nominal operation. High DREI values indicate configurations that maximize system utility while minimizing transition cost under evolving threat conditions.

\section{Methodology}\label{sec:simulation_setup}

\subsection{Simulation Environment}\label{sec:simulation_environment}

To evaluate the proposed framework, we simulate a representative signal jamming scenario using a time-stepped, event-driven model. The satellite system evolves under operational constraints and external threats as defined in Fig.~\ref{fig:satellite_system_diagram}. At each timestep, the environment updates energy reserves, jamming intensity \(J_t \in [0,1]\), and threat activity, then applies fallback transitions based on the selected PACE model. Each transition incurs a probabilistic cost and alters the system state. A predefined crisis at a fixed timestep intensifies threats and tests recovery via layer suppression, blocked transitions, or elevated transition costs.

\subsection{Implemented PACE Variants}\label{sec:pace_variants}

We implement three operational models of the PACE framework, each operating on the same underlying layered graph structure but differing in how transitions are selected at runtime.

\subsubsection{Simple Static PACE Model}\label{sec:simple_pace}

This baseline uses fixed transition weights from precomputed threat analyses, making it functionally analogous to conventional static redundancy schemes such as N+1 or TMR, which apply predetermined responses without runtime adaptation. Each transition \((s, s') \in \mathcal{T}\) is defined by:
\begin{equation}
w(s, s') = \left( p(s, s'),\; c(s, s') \right)
\end{equation}

At each timestep, the system stays in state \(s\) with probability \(p_{\text{stay}}\); otherwise, it transitions according to the static distribution \(\{p(s, s')\}\). This non-reactive model ignores environmental feedback and serves as the primary baseline.

\subsubsection{Environment-Aware Adaptive PACE Model}\label{sec:adaptive_pace}

This model adjusts transitions based on real-time conditions, specifically jamming intensity \(J_t\) and recovery concurrency \(C \geq 1\). For each transition \((s, s')\), parameters evolve as:
\begin{align}
p_t(s, s') &= p(s, s') \cdot (1 + \lambda J_t) \\
c_t(s, s') &= c(s, s') + \gamma J_t - \alpha (C - 1),
\end{align}
with sensitivity coefficients \(\lambda, \gamma, \alpha\). Higher \(J_t\) increases the likelihood of degraded-state transitions, while concurrent recovery lowers cost. After scaling, probabilities are clipped to \([0,1]\) and renormalized to form a valid distribution; costs are clipped to a minimum of zero to prevent infeasible negative-cost transitions.

\subsubsection{Epsilon-Greedy MDP-Inspired PACE Model}\label{sec:greedy_pace}

This model applies a one-step reward-based decision rule inspired by Markov Decision Processes. For each valid transition \((s, s')\), the reward is:
\begin{equation}
R(s, s') = \omega(s') - c_t(s, s'),
\end{equation}
where \(\omega(s')\) is the utility of the target state and \(c_t(s, s')\) is the transition cost adjusted for current environmental conditions. Action selection follows an \(\epsilon\)-greedy policy: with probability \(\epsilon\) a random valid transition is chosen (exploration), and with probability \(1 - \epsilon\) the highest-reward transition is selected (exploitation):
\begin{equation}
s \rightarrow s' \sim
\begin{cases}
\text{Uniform}(\mathcal{A}(s)) & \text{with probability } \epsilon \\
\arg\max_{s' \in \mathcal{A}(s)} R(s, s') & \text{with probability } 1 - \epsilon,
\end{cases}
\end{equation}
where \(\mathcal{A}(s)\) denotes the set of valid successor states. This approach injects randomness to avoid local optima while favoring transitions that maximize utility relative to cost.

For each PACE variant, we perform 5000 independent Monte Carlo trials, each running up to \(T = 12\) timesteps.

\section{Simulation Results and Discussion}\label{sec:results_discussion}

\subsection{Comparative Performance of PACE Models}

\begin{figure}[!t]
\centering
\includegraphics[width=\linewidth]{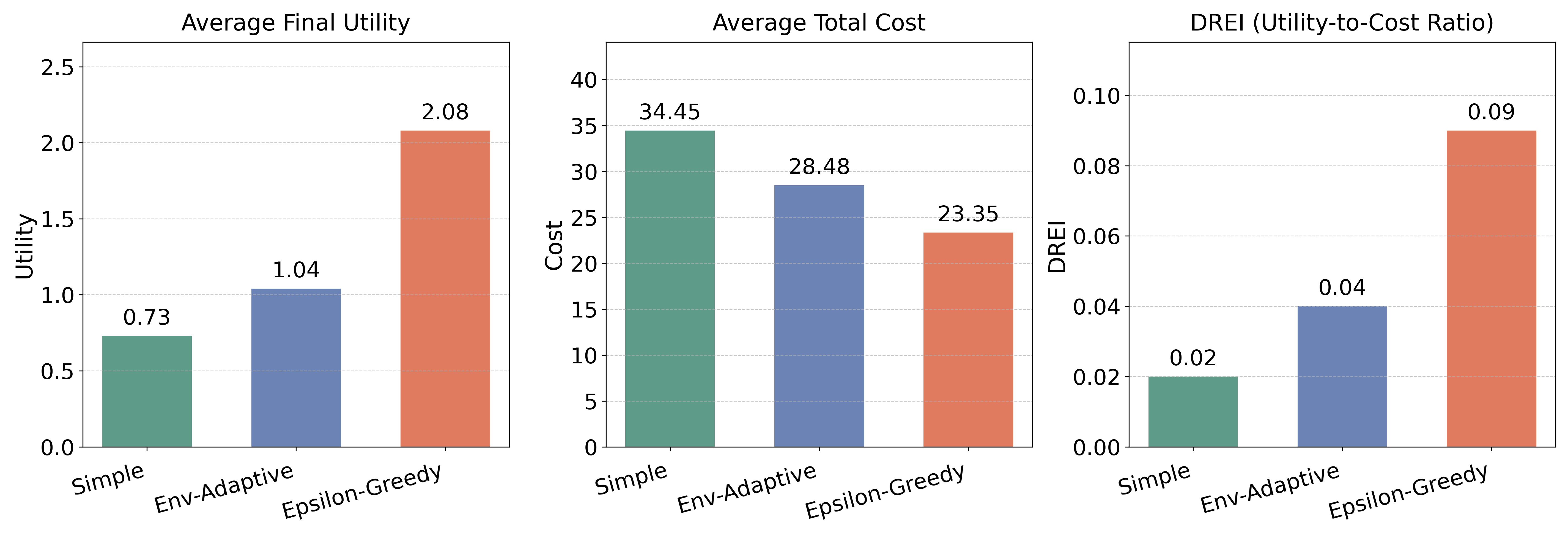}
\caption{Average final utility (left), total cumulative cost (center), and DREI (right) across all trials for the static, adaptive, and \(\epsilon\)-greedy PACE models.}
\label{fig:avg_bar_plots}
\end{figure}

\begin{figure}[!t]
\centering
\includegraphics[width=\columnwidth]{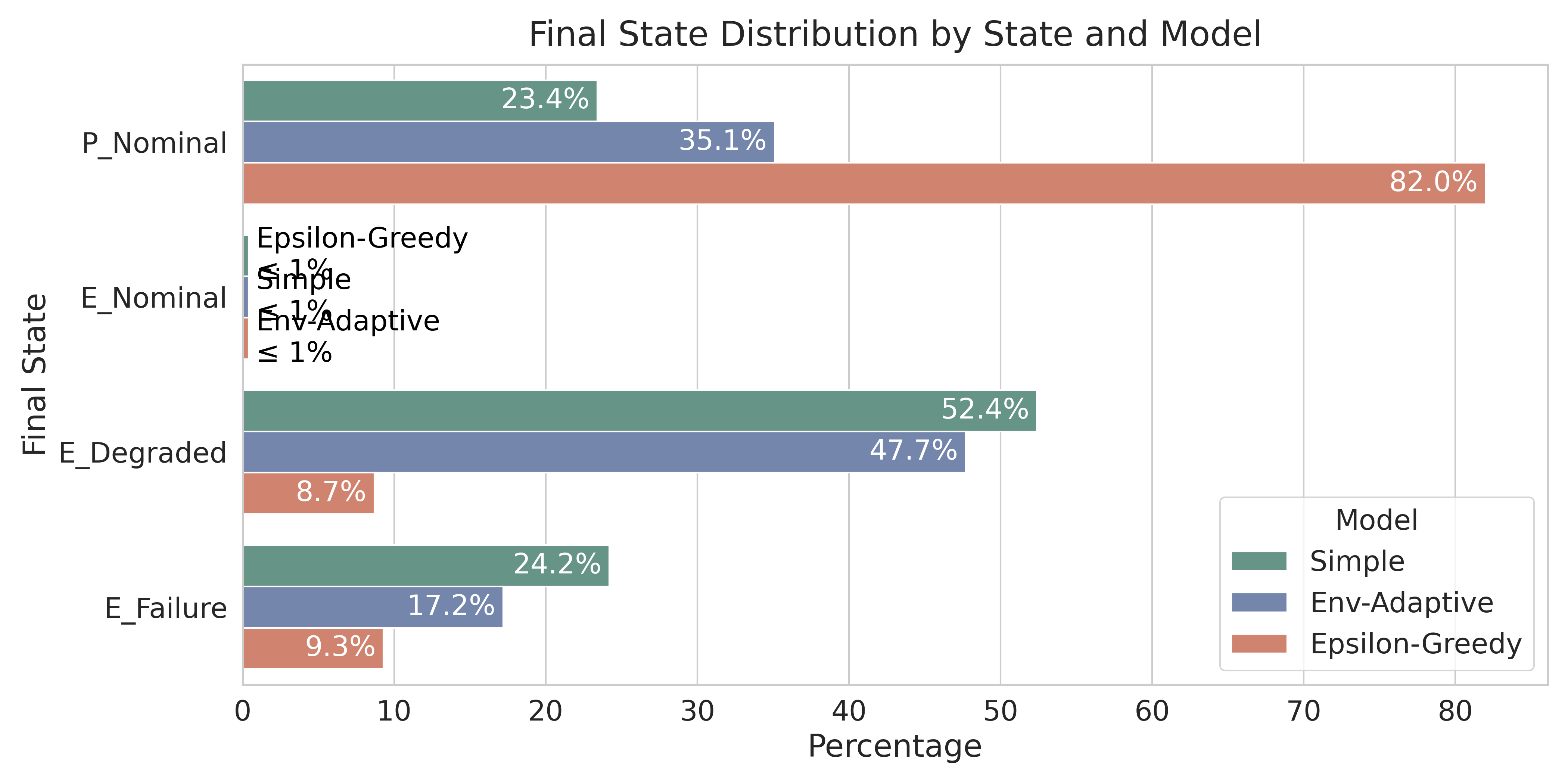}
\caption{Final operational state distributions showing the proportion of runs ending in nominal, degraded, or failure states across models.}
\label{fig:final_state_dist}
\end{figure}

Fig.~\ref{fig:avg_bar_plots} summarizes average final utility, total cost, and DREI across 5000 Monte Carlo trials. The static model achieved 0.73 utility, 34.45 cost, and 0.02 DREI. The adaptive model improved to 1.04 utility, 28.48 cost, and 0.04 DREI. The \(\epsilon\)-greedy model outperformed both, with 2.08 utility, 23.35 cost, and 0.09 DREI, demonstrating superior balance across all three metrics.

Fig.~\ref{fig:final_state_dist} presents final state distributions. The static model resulted in 24.2\% failures, 52.4\% degraded, and 23.4\% nominal states. The adaptive model reduced failures to 17.2\% and increased nominal states to 35.1\%. The \(\epsilon\)-greedy model was most resilient, achieving 82.0\% nominal, 8.7\% degraded, and 9.3\% failure. The cost histograms in Fig.~\ref{fig:combined_cost_hist_cdf} confirm that these differences are consistent across trials: the static model shows broad, high-variance cost distributions while the \(\epsilon\)-greedy model produces a tight, low-cost cluster, indicating that the performance gap is robust rather than driven by outliers.

\subsection{Temporal Dynamics of Utility and Cost}

\begin{figure}[!t]
\centering
\begin{subfigure}[b]{\linewidth}
    \includegraphics[width=0.97\linewidth]{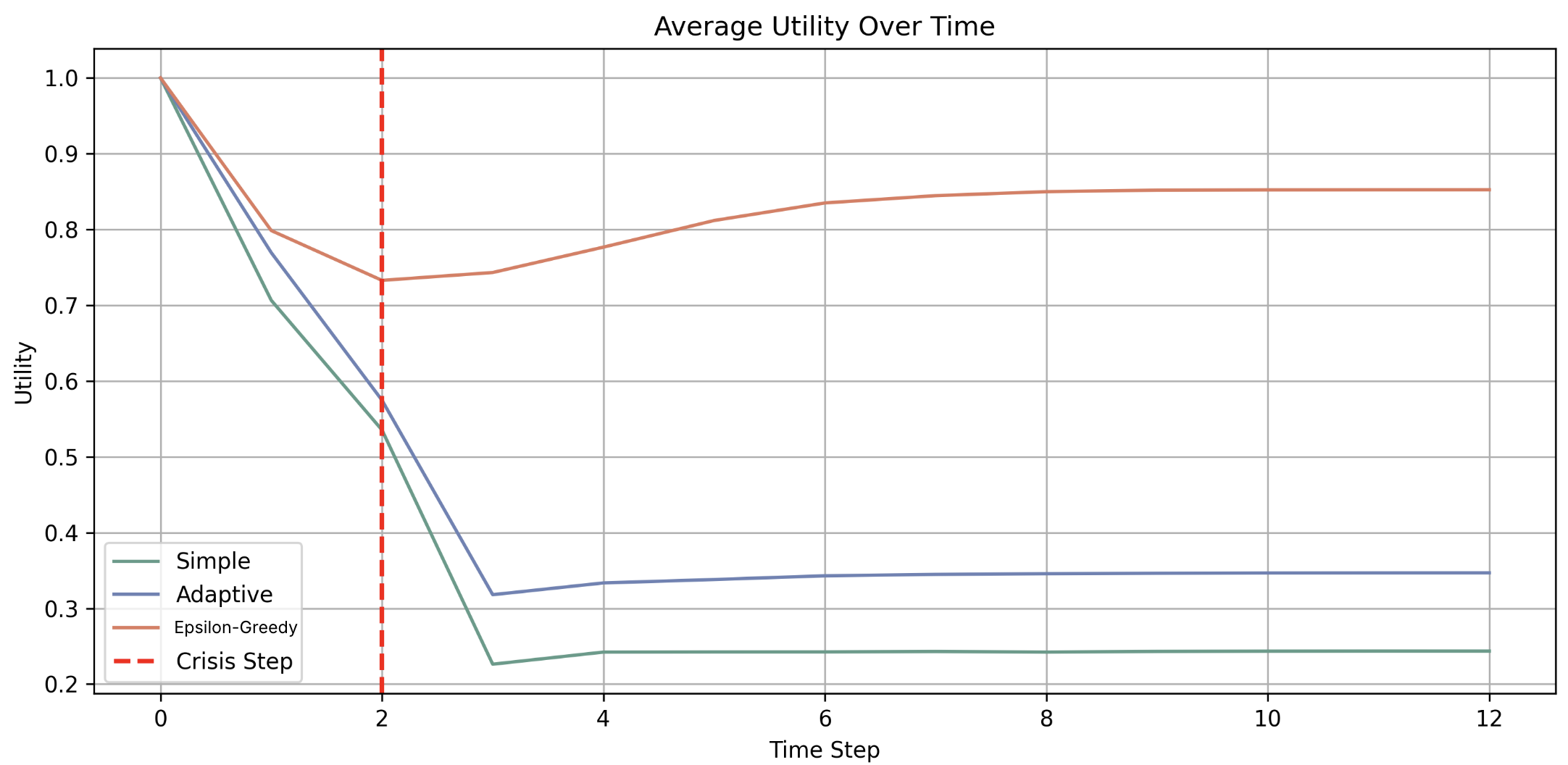}
    \caption{Average utility over time.}
    \label{fig:avg_utility_over_time}
\end{subfigure}
\begin{subfigure}[b]{\linewidth}
    \includegraphics[width=0.97\linewidth]{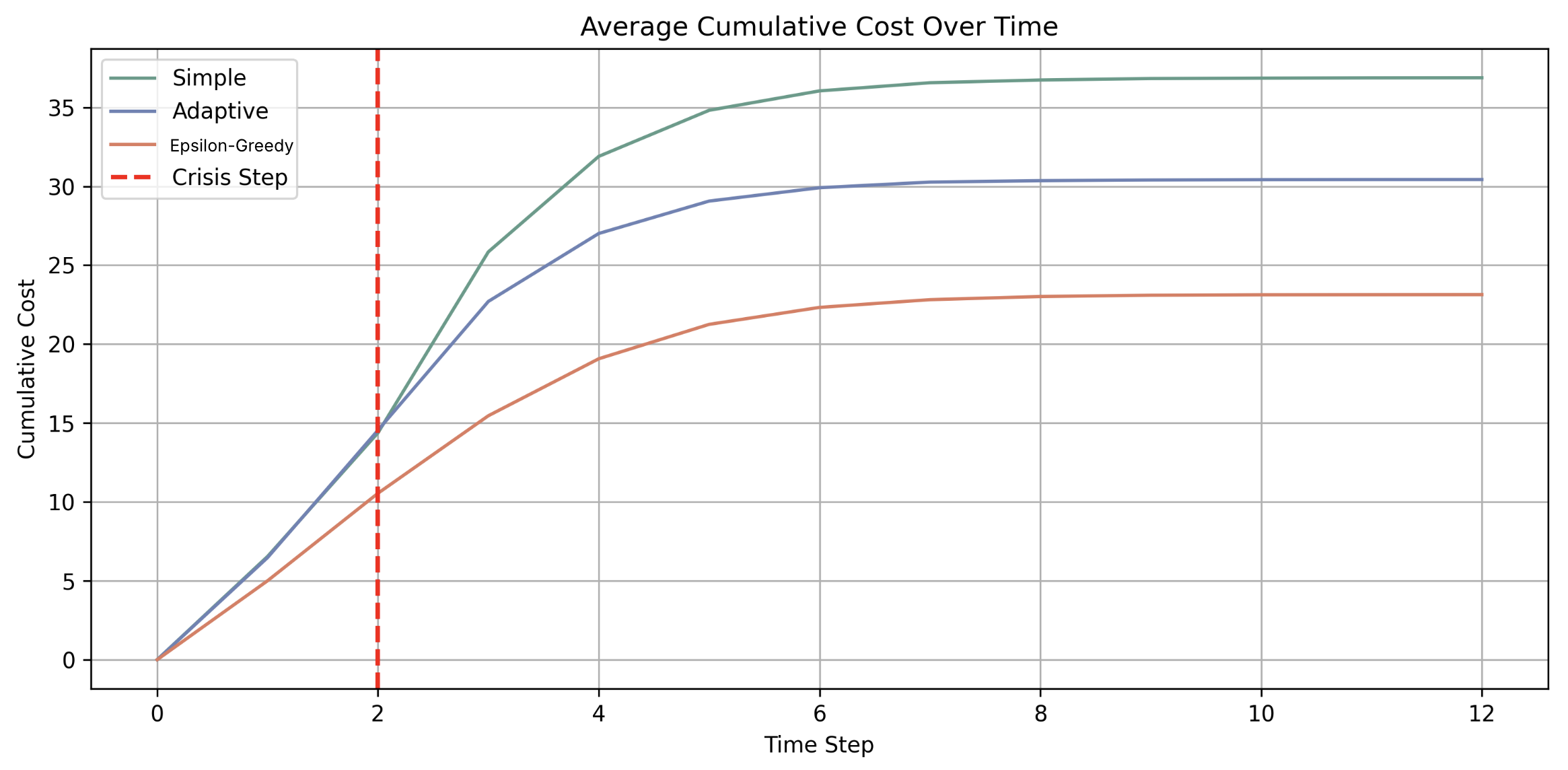}
    \caption{Average cumulative cost over time.}
    \label{fig:avg_cost_over_time}
\end{subfigure}
\caption{Temporal evolution of utility and cost. The red dashed line marks the crisis event at timestep 2.}
\label{fig:utility_cost_over_time}
\end{figure}

Fig.~\ref{fig:utility_cost_over_time}(a) shows average utility over time. Following the crisis at timestep 2, utility dropped sharply for the static (to 0.22) and adaptive (to 0.32) models. The \(\epsilon\)-greedy model recovered steadily, stabilizing near 0.85 at the end of the horizon.

Fig.~\ref{fig:utility_cost_over_time}(b) shows cumulative cost. The static model rapidly accrued costs, saturating near 37 units. The adaptive model leveled off near 30, while the \(\epsilon\)-greedy model rose slowly, reaching just 23 units. The \(\epsilon\)-greedy model's post-crisis recovery indicates greater adaptability in restoring performance, and its lower cost accumulation reflects improved efficiency in fallback transitions.

\subsection{Cost Distributions and Variability}

\begin{figure}[!t]
\centering
\includegraphics[width=\columnwidth]{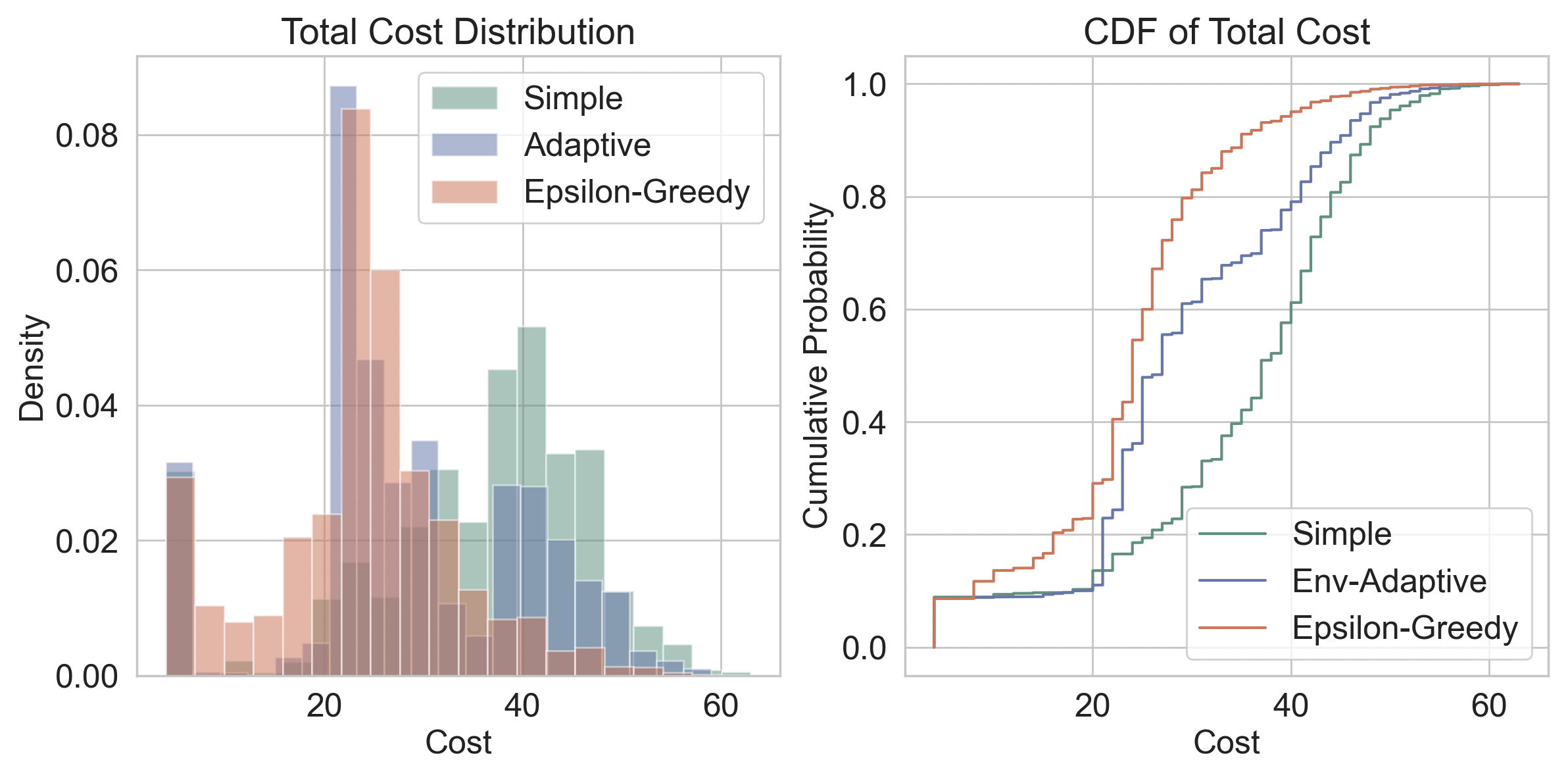}
\caption{Total cost distribution (left) and cumulative distribution function (right) across models.}
\label{fig:combined_cost_hist_cdf}
\end{figure}

Fig.~\ref{fig:combined_cost_hist_cdf} summarizes total cost distributions. The static model's cost distribution is broad and skewed toward higher values, reflecting large outcome variability. The adaptive model's distribution is narrower, with a central mode around 30--35 units. The \(\epsilon\)-greedy model produces the most compact distribution, concentrated around 20--25 units, indicating consistent low-cost outcomes. The CDF confirms these trends: the \(\epsilon\)-greedy curve rises steeply, the adaptive model follows, and the static model exhibits the slowest slope and highest variability.

\subsection{Key Insights and Interpretation}

The results highlight the value of incorporating environmental feedback into fallback strategies. While the static model struggled with dynamic disruptions, the adaptive version showed modest gains by reacting to threat levels. The \(\epsilon\)-greedy model was notably more resilient, recovering utility more effectively while keeping costs lower.

Regarding hyperparameter robustness, we observed qualitatively stable rankings across tested ranges: \(\epsilon \in [0.05, 0.3]\), \(\lambda \in [0.3, 0.7]\), \(\gamma \in [0.1, 0.4]\), \(\alpha \in [0.05, 0.2]\), and \(p_{\text{stay}} \in [0.4, 0.7]\). Increasing \(\epsilon\) reduced exploitation efficiency but improved recovery in corner cases; increasing \(\lambda\) accelerated downward transitions under high jamming, consistent with the model's design intent. The DREI ranking remained stable across these ranges, confirming that the \(\epsilon\)-greedy model's advantage is not confined to a narrow parameter regime.

The numerical evaluation focuses on a single jamming scenario over a short horizon (\(T = 12\)), which, while sufficient to demonstrate the relative merits of the three PACE variants, does not fully validate the broader multi-threat, multi-segment resilience claims made in the threat taxonomy. Evaluating sustained or multi-crisis jamming, as well as cyber-intrusion and EMP scenarios, remains future work. Capturing longer-term dependencies, such as how sustained jamming affects power consumption or recovery feasibility, would further improve realism.

\section{Conclusion}\label{sec:conclusion}

Our comparative analysis across three PACE variants (static, adaptive, and \(\epsilon\)-greedy reward-optimized) demonstrated that traditional static redundancy fails to maintain utility under crisis conditions, often collapsing into degraded or failed states. The adaptive model improved resilience by incorporating environmental feedback, yet still showed limitations under high-stress scenarios. The \(\epsilon\)-greedy model consistently outperformed both baselines, achieving the highest recovery rates, the lowest cumulative costs, and the highest DREI. These results highlight several critical insights for satellite system design. First, resilience cannot be delivered through static redundancy alone; systems must integrate real-time adaptation to dynamic threats. Second, lightweight decision mechanisms such as reward-based \(\epsilon\)-greedy policies can deliver substantial resilience gains without requiring full reinforcement learning or high computational overhead. Third, designing for graceful degradation across multiple operational layers not only improves mission continuity but also reduces the risk of catastrophic failure. Future work will extend the validation to cyber-intrusion and EMP threat scenarios, longer simulation horizons, and multi-crisis sequences to substantiate the framework's broader applicability.

\section*{Acknowledgment}

This work was supported in part by the Tier 1 Canada Research Chair Program. The authors thank the team at NorthStar Earth \& Space, including Narendra Gollu, Naron Phou, Noemi Giammichele, Yann Picard, Jean-Claude Leclerc, and Srinivas Setty, for their discussions, technical insights, and collaborative support.

\bibliographystyle{IEEEtran}
\bibliography{reference}

\end{document}